\documentclass[a4paper,10pt]{article}
\usepackage{graphicx}

\begin{document}

\setlength{\oddsidemargin} {1cm}
\setlength{\textwidth}{18cm}
\setlength{\textheight}{23cm}

\title{Inhomogeneities in the Universe with exact solutions of General Relativity} 
\author{{Marie-No\"elle C\'el\'erier} \\ 
{\small Laboratoire Univers et TH\'eories (LUTH), Observatoire de Paris, CNRS \& Universit\'e Paris Diderot} \\
{\small 5 place Jules Janssen, 92190 Meudon, France} \\
{\small e-mail: marie-noelle.celerier@obspm.fr}}

\maketitle

\begin{abstract}

It is commonly stated that we have entered the era of precision cosmology in which a number of important observations have reached a degree of precision, and a level of agreement with theory, that is comparable with many Earth-based physics experiments. One of the consequences is the need to examine at what point our usual, well-worn assumption of homogeneity associated to the use of perturbation theory begins to compromise the accuracy of our models. It is now a widely accepted fact that the effect of the inhomogeneities observed in the Universe cannot be ignored when one wants to construct an accurate cosmological model. Well-established physics can explain several of the observed phenomena without introducing highly speculative elements, like dark matter, dark energy, exponential expansion at densities never attained in any experiment (i.e. inflation), and the like. Two main classes of methods are currently used to deal with these issues. Averaging, sometimes linked to fitting procedures a la Stoegger and Ellis, provide us with one promising way of solving the problem. Another approach is the use of exact inhomogeneous solutions of General Relativity. This will be developed here.
 
\end{abstract}

\section{Introduction}
\label{int}

The observation of its structures show that our Universe is not homogeneous. We see voids, groups of galaxies, clusters, superclusters, walls, filaments, etc. However, it is usually argued in the literature that the Universe should be nearly homogeneous at large scales which is supposed to validate the use of Friedmannian models. But how large these scales are and what nearly does imply is never precisely stated.

It has however become, during the last few years, as a widely accepted fact, that the effect of the inhomogeneities cannot be ignored when one wants to construct an accurate cosmological model up to the regions where structures start forming and their evolution becomes non-linear. Three different methods have been proposed to deal with this issue:

\begin{enumerate}

\item 
Linear perturbation theory. However, this method is only valid when {\it both} the curvature and density contrasts remain small, which is not the case in the non-linear regime of structure formation and where the SNe Ia are observed.

\item 
Averaging methods `{\`a} la Buchert', promising, but needing to be improved (see \cite{TB08} and references therein).

\item
Exact inhomogeneous solutions, valid at all scales and exact perturbations of the Friedmann background which they can reproduce as a limit with any precision.

\end{enumerate}

The use of such exact solutions shows that well-established physics can explain several of the phenomena observed in astrophysics and cosmology without introducing highly speculative elements, like dark matter, dark energy, exponential expansion at densities never attained in any experiment (inflation), and the like. Here, we foccuss on the application of a couple of exact solutions of general relativity to structure formation and evolution and to the reproduction of cosmological data.

\section{Use of exact inhomogeneous models in astrophysics and cosmology}
\label{eim}

Very few exact inhomogeneous solutions of Einstein's equations have been used for these purposes. Those which appear most often are:

\begin{enumerate}

\item
The Lema\^itre -- Tolman (L--T) models \cite{GL33,RT34} which are spherically symmetric dust solutions of Einstein's equations. They are determined by one coordinate choice and two free functions of the radial coordinate $r$ chosen among three independent ones: the energy per unit mass , $E(r)$, of the particles contained within the comoving spherical shell at a given $r$, the gravitational mass, $M(r)$, contained in that shell and the Bang time function, $t_B(r)$, meaning that the Big Bang occurred at different times at different $r$ values. The homogeneous FLRW model is one sub-case.

\item
The Lema\^itre model \cite{GL33} (usually known as Misner-Sharp \cite{MS64}) is not an explicit solution but a metric determined by a set of two differential equations. It represents a spherically symmetric perfect fluid with pressure gradient. Its solution is obtained by numerical integration.

\item
The Quasi-spherical Szekeres (QSS) models \cite{PS75} are dust solutions of Einstein's equations with no symmetry at all. They are defined by one coordinate choice and five free functions of the radial coordinate. The L--T and FLRW models are sub-cases.

\item
The spherically symmetric Stephani models \cite{HS67} have also been used for cosmological purpose. They are exact solutions with homogeneous-energy density and inhomogeneous-pressure.

\end{enumerate}

L--T models have been the most widely used in cosmology since they are the most tractable among the few available ones cited above. However, QSS models are currently slightly coming into play.

But caution with L--T models is required since:

\begin{itemize}

\item  An origin, or centre of spherical symmetry, occurs at $r = r_c$ where $R(t,r_c)= 0$ for all $t$ (here, $R$ is the areal radius). The conditions for a regular centre were derived by Mustapha and Hellaby \cite{MH01}.

\item  Shell crossings, where a constant $r$ shell collides with its neighbour, create undesirable singularities while the density diverges and changes sign. The conditions on the 3 arbitrary functions of the model that ensure none be present anywhere in an L--T model are given in \cite{HL85}.

\item  The assumption of central observer, generally retained for simplicity, can be considered as grounded on the observed quasi-isotropy of the CMB temperature, and thus as a good working approximation at large scales. At smaller scales, it gives simplified models of the Universe averaged over the angular coordinates around the observer, i. e., with the relax of only one degree of symmetry as regards the homogeneity assumption. The central observer location is just an artefact of this angular averaging.

\end{itemize}

However, models assuming a non-central observer and L--T Swiss-cheeses have also been studied to get rid of possible misleading features of spherical symmetry.

\section{Structure evolution}

\subsection{Structure evolution with L--T and Lema\^itre models}

The L--T class of solutions have been used to reproduce the formation and evolution of structures from the near homogeneity seen in the CMB temperature.

In \cite{KH04}, the evolution from an initial density profile to a galaxy cluster whose density profile approximate the `Universal Profile' of an Abell cluster implies a density amplitude at the initial time which differs from the observed values at last scattering by three orders of magnitude. However, the velocity amplitude at the initial time obtained by the same process is on the border of the observationally implied range. This shows that velocity perturbations generate structures much more efficiently than density perturbations. Moreover, it is demonstrated in \cite{MH01} that a smooth evolution can take an initial condensation to a void. This implies that the initial density distribution does not determine the final structure which will emerge from it. The velocity distribution can obliterate the initial setup.

However, a void consistent with the observational data (density contrast less than $\delta = -0.94$, smooth edges and high density in the surrounding regions) is very hard to obtain with L--T models without shell crossing \cite{BKH05}.

But adding a realistic distribution of radiation (using Tolman models) helps forming such voids \cite{KB06a}.

\subsection{Structure evolution with quasi-spherical Szekeres models}

In the works reported in the above subsection, the spherically symmetric L--T and Lema\^itre models were used to study structure formation. However, the structures observed in the Universe are far from being spherical and the analysis needs thus to be refined by considering a wider variety of astrophysical objects with different shapes. The investigation of the evolution of small voids inside compact clusters and of large voids surrounded by walls or filaments has been performed with QSS models.

A void with an adjourning supercluster evolved inside an homogeneous background (FIGURE 1) has been considered in \cite{KB06b}. To estimate how two neighbouring structures influence each other, the evolution of a double structure in QSS models (FIGURE 2) has been compared with that of a single structure in L--T models and in linear perturbation theory. In the QSS models studied, the growth of the density contrast is 5 times faster than in the corresponding L--T models and 8 times faster than in the linear approach. This could imply a strong improvement of the phenomenon of structure formation which, in the standard CDM model, is too slow to form the structures corresponding to the observations made at larger and larger redshifts. However, the evolution of the void is slower within the Szekeres model than it is in the L--T model. This suggests that single, isolated voids evolve much faster than the ones which are in the neighbourhood of large overdensities where the mass of the perturbed region is above the background mass.

\begin{figure}
  \includegraphics[height=.3\textheight, scale=0.5]{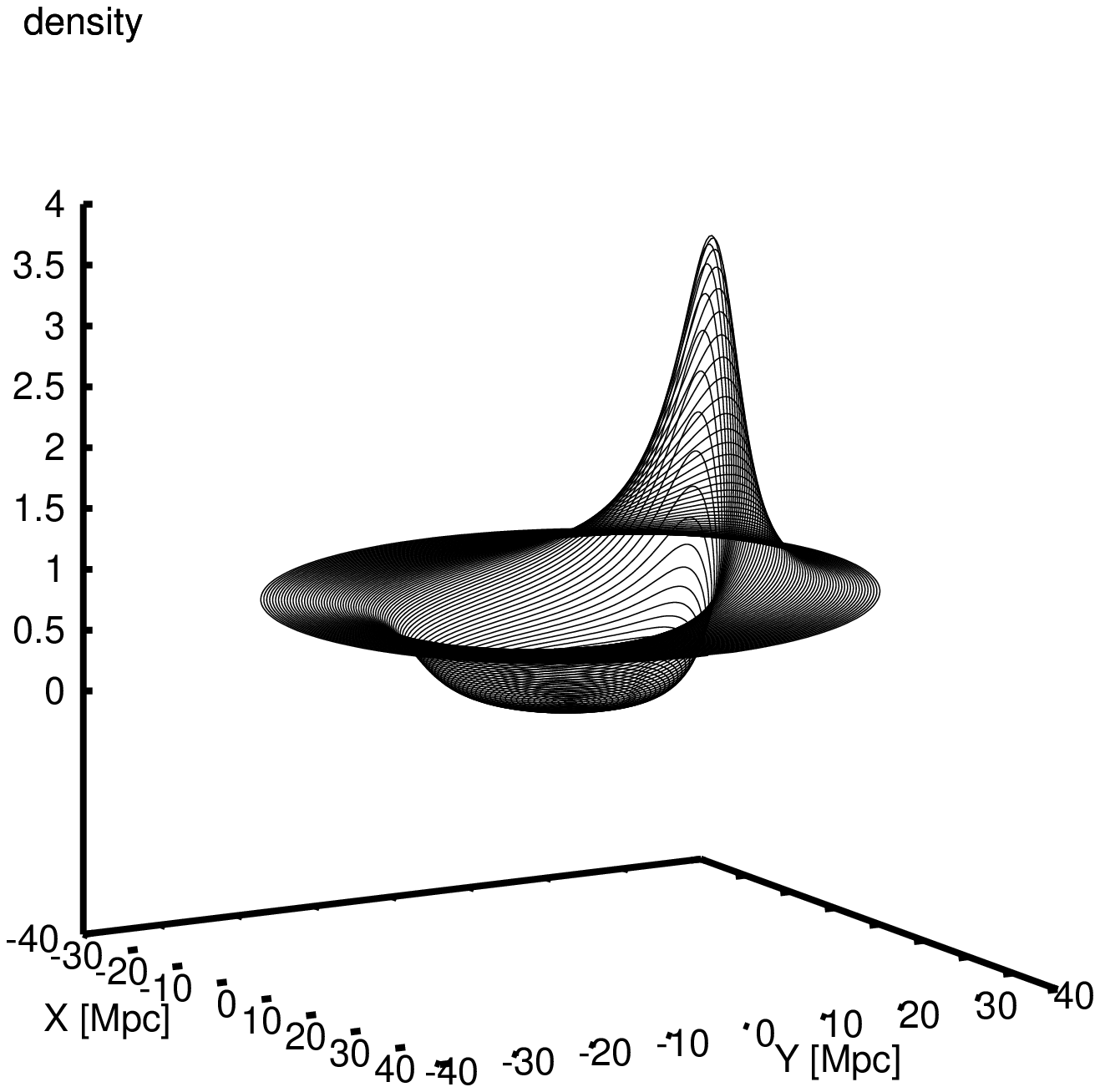}
  \includegraphics[height=.3\textheight, scale=0.5]{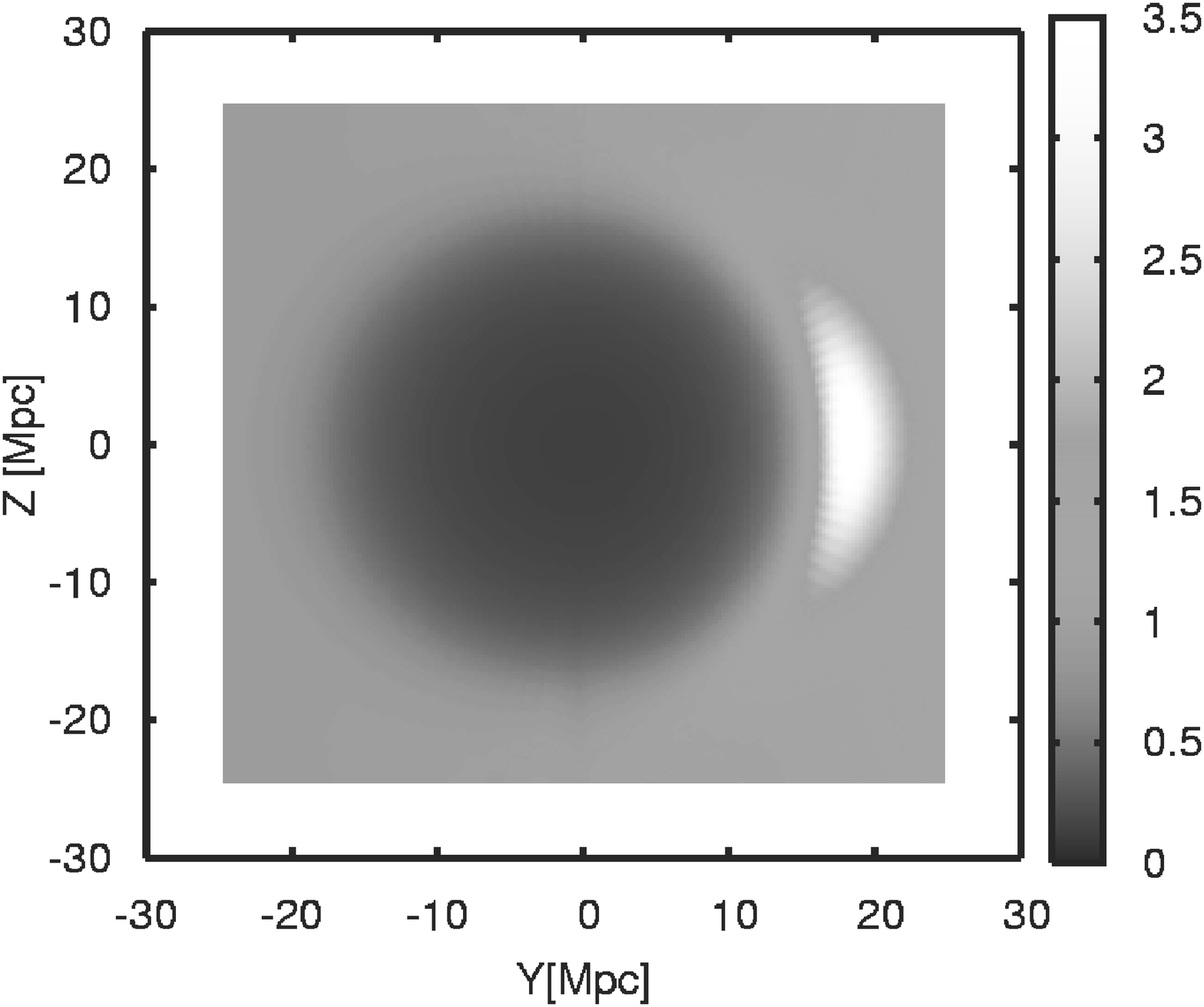} 
  \caption{Double structure: the present-day density distribution, $\rho/\rho_b$, of a double-structure model in back-ground units. The left panel shows the iso-density curves in the spatial section X-Y. The right panel shows the present day colour-coded density distribution in the spatial section Y-Z. White represents high-density and black, low-density regions}.
\end{figure}

\begin{figure}
  \includegraphics[height=.3\textheight]{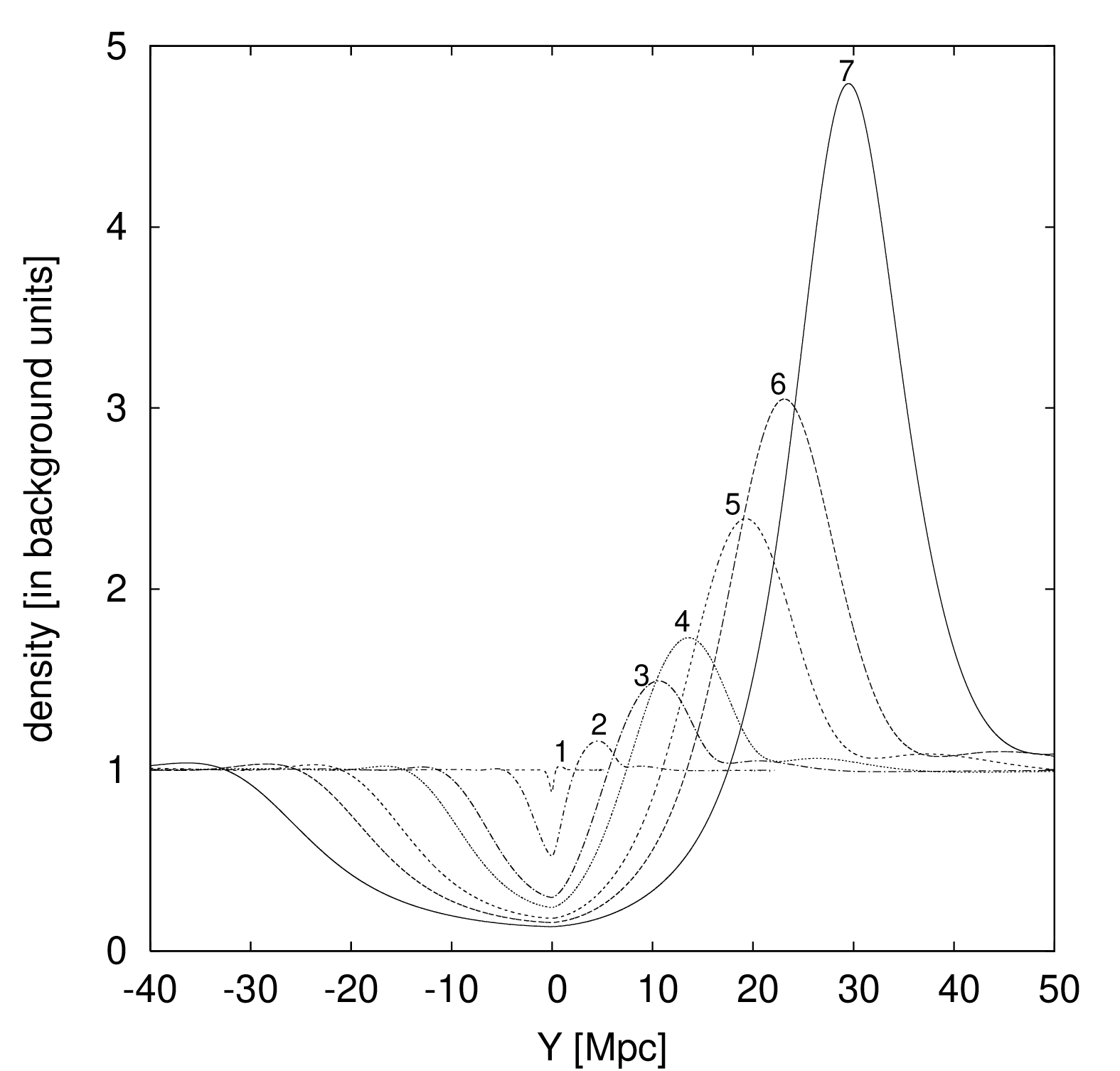}
  \caption{Double structure: evolution of the density profile from 100 Myr after the Big Bang (1) up to the present time (7).}
\end{figure}

The model of triple structure is composed of an overdense region at the origin, followed by a small void which spreads to a given $r$ coordinate. At a larger distance from the origin, the void is huge and its larger side is adjacent to an overdense region \cite{KB07} (FIGURE 3). Where the void is large, it evolves much faster than the underdense region closer to the `centered' cluster. The exterior overdense region close to the void along a large area evolves much faster than the more compact supercluster at the centre (FIGURE 4). This confirms that, in the Universe, small voids surrounded by large high densities evolve much more slowly than large isolated voids.

\begin{figure}
  \includegraphics[height=.3\textheight]{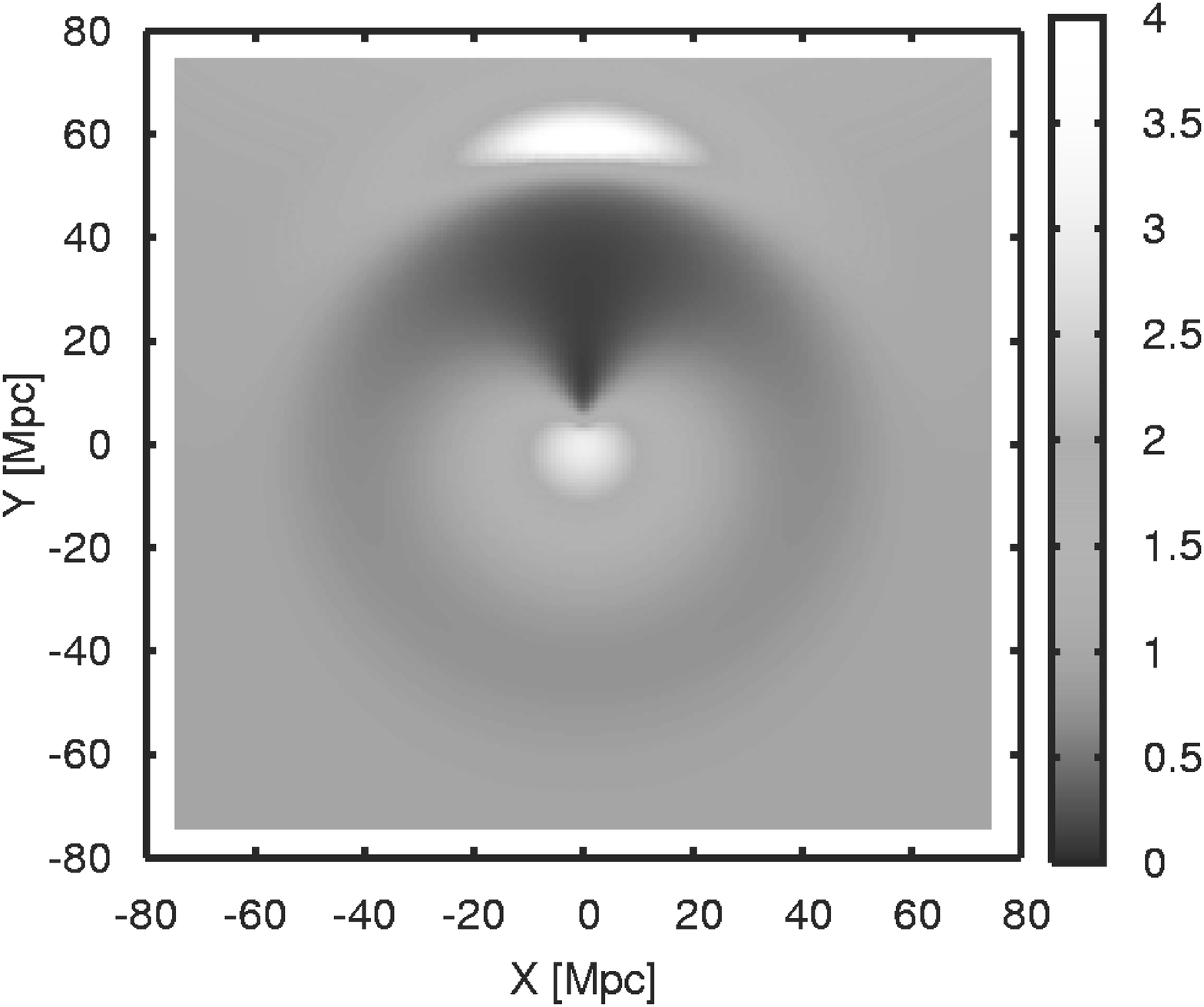}
  \caption{Triple structure: the present-day colour-coded density distribution, $\rho/\rho_b$, of a triple-structure model in back-ground units, showing a slice through the origin.}
\end{figure}

\begin{figure}
  \includegraphics[height=.4\textheight, angle=270]{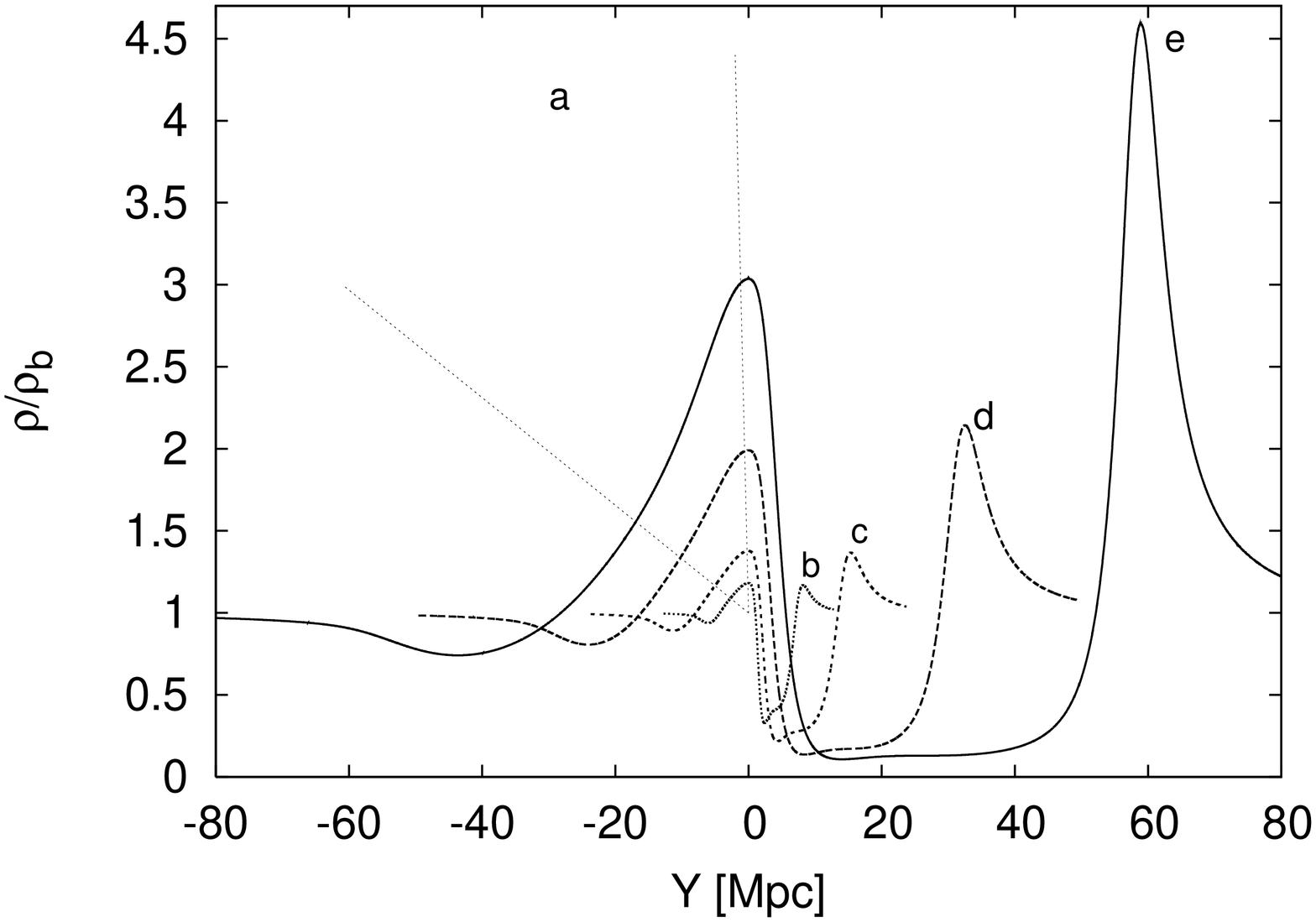}
  \caption{Triple structure: evolution of the density profile from 0.5 Myr after the Big Bang (a) up to the present time (e).}
\end{figure}

\section{The dark energy puzzle}

Since its discovery during the late 1990s \cite{Ri98,Pe99}, the `dimming' of distant type Ia supernovae has been mostly ascribed to the influence of a mysterious dark energy component, i.e. an unknown `fluid' or `field' with negative pressure. Formulated in a Friedmannian framework, based upon the `cosmological principle', this interpretation has given rise to the `Concordance' model where the Universe expansion is accelerated by the dark energy pressure.

However, what we observe is {\it not an accelerated expansion} (this is only the outcome of the Friedmannian assumption) but the dimming of the supernovae as regards their luminosity predicted in the previous Einstein-de Sitter standard model of cosmology. More exactly we establish their luminosity distance-redshift relation, itself inferred from the flux measurement of their light curves.

Shortly after this discovery, it was proposed by a small number of authors that this effect could be due to the large-scale inhomogeneities of the Universe \cite{PS99,MNC00,KT00}. After a period of relative disaffection, this proposal experienced a renewed interest about five years ago.

Now, the accelerated expansion interpretation was sufficiently misleading such as to induce some authors to try to derive or rule out no-go theorems, i.e., theorems stating that a locally defined expansion cannot be accelerating in inhomogeneous models satisfying the strong energy condition. But, as we have seen above, this is not the point.

Other authors stressed, more accurately, that the definition of a deceleration parameter in an inhomogeneous model is tricky \cite{HS05,Ap06} and has nothing to do with reproducing the supernova data \cite{KHCB09}.

It is well-known, from the work of Mustapha, Hellaby and Ellis \cite{MHE97} that an infinite class of L--T models can fit a given set of observations isotropic around the observer. This has been used by C\'el\'erier \cite{MNC00} to examplify her demonstration that models which are spherically symmetric around the observer can fit the supernova data and that the problem is completely degenerate. This is the reason why many different central observer L--T models have been proposed in the literature and shown to succeed rather well.

Thus, to constrain the model further on, it is mandatory to fit it to other cosmological data.

\subsection{The direct and inverse problems}

There exist two procedures for trying to explain away dark energy with (L--T) models:

\begin{enumerate}

\item
The direct way uses a smaller number of degrees of freedom than allowed. Here, one first guesses the form of the parameter functions defining a class of models supposed to represent our Universe with no cosmological constant or dark energy, and writes the dependence of these functions in
terms of a limited number of constant parameters. Then one fits these
constant parameters to the observed SN Ia data or to the luminosity distance-redshift relation of the $\Lambda$CDM model.

\item
The inverse problem is more general. It amounts to consider the luminosity distance $D_L(z)$ as given by observations or by the $ \Lambda$CDM model as an input and try to select a specific L--T model with zero cosmological constant best fitting this relation.

\end{enumerate}

Then, to avoid degeneracy, one must jump to a further step and try to reproduce more and possibly all the available observational data, \cite{AA06,KB08,GBH08,LH07,MCH07,HA08}.

\subsection{Matching observations with a single patch L--T model exhibiting a central void}

The L--T solutions with a central observer has been used as a first step in the process of reproducing cosmological inhomogeneities. They model the Universe with the inhomogeneities smoothed out over angles around the observer whose location can be anywhere (this does not contradict any Copernican principle). This is analogous to the smoothing out over the whole space in homogeneous models. The use of such models must be regarded as a first approach which will be followed in the future by more precise ways of dealing with the observed inhomogeneities.

As it has been recalled in the second section of this article, an L--T model is defined by two independent arbitrary functions of the radial coordinate, which must be fitted to the observational data. However, in most of the proposals available in the literature, the generality of the models have been artificially limited by giving the initial-data functions a handpicked algebraic form depending on the authors' feelings about which kind of model would best represent our Universe. Only a few constant parameters are left arbitrary to be adapted to the observations.

Another way in which the generality of the L--T models has been artificially limited is the assumption that the age of the Universe is everywhere the same, i.e. that the L--T bang-time function $t_B$ is constant. With $t_B$ being constant, the only single-patch L--T model that fits observations is one with a giant void \cite{INN02,YKN08}. The argument brought in defense of the constant $t_B$ assumption is that a
non-constant $t_B$ generates decreasing modes of perturbation of the metric \cite{JS77,PK06}, so any substantial inhomogeneity at the present time stemming from $t_{B,r} \neq 0$ would imply huge perturbations of homogeneity at the last scattering. This, in turn, would contradict the CMB observations and the implications of inflationary models.

In the recent years, we have thus seen the increase in popularity of a large, then huge, then giant void model, where the observer is located at or near the centre of a large, huge, giant L--T void of size of up to
a few Gpc. The most achieved example of this class of models is the GBH model, specified by its matter content $\Omega_M(r)$ and its expansion rate $H(r)$, governed by 5 free constant parameters and  matching to an Einstein-de Sitter universe at large scales \cite{GBH08}. It is fitted to a series of observations (CMB, LSS, BAO, SN Ia, HST measure of $H_0$, age of the globular clusters, gas fraction in clusters, kinematic Sunyaev-Zel'dovich effect for 9 distant galaxy clusters) used to constrain its parameters. With this class of models, the conclusion is that the possibility that we leave close to the centre of a large (around 2.5 Gpc) void within an Einstein-de Sitter Universe with no dark energy is not excluded.

However, as many other central void models, the GBH model is constructed with a central underdensity defined a priori by a set of constant parameters (the underdensity at the centre of the void, the size of the void and the transition width of the void profile). Hence, the outcome of the fit can be nothing but a central void.

Why, then, several researchers have been led astray by the frequent claims of a giant void being implied by an L--T model? The reason might be twofold.

From the results of an analysis of 44 SN Ia distances they performed in the framework of a FLRW model with $\Omega_m = 1$ and $\Omega_{\Lambda} = 0$, Zehavi {\em et al.} \cite{IZ98} suggested a monopole in the peculiar velocity field they interpreted as marginal evidence for a local void of radius 70 $h^{-1}$ Mpc and $\sim $20\% underdensity surrounded by a dense shell roughly coinciding with the local great walls. They called this putative local void the `Hubble Bubble',
and, under this name, it became popular in the literature.

Strangely, a mere toy model can become nearly as strong a reference as actual observations. In a series of articles, Tomita \cite{KT00} used a simple toy model to explain in a very pedagogical way how `dark energy' could in principle be mimicked by a local void. He assumed a low-density inner homogeneous region is connected at some redshift to an outer homogeneous region of higher density. Both regions decelerate, but since the inner void expands faster than the outer region, an apparent acceleration is experienced by the observer located inside this void and looking at supernovae bursting in the outer region. Such a toy model was only designed to stress that an actual accelerated expansion is not needed to reproduce the SN Ia data. However, many authors have taken this example literally and cited this model as if it were evidence for the existence of a local void.

\subsection{We can do without the giant void in a central observer L--T model}

Since, as we show below, the arguments in favour of the limitations imposed above on the arbitrary functions of the radial coordinate lack sufficient strength, there is no need to restrict a priori their degrees of freedom.

The arguments in favour of or against a local `Hubble Bubble' have been disputed at length within the astronomical community and the latest results seem to point to our being located in a `Local Sheet' ($\sim$ 7 Mpc long) which bounds a `Local Void', a nearly empty region with radius at least 23 Mpc \cite{TS08,JR09}. This seems to contradict the putative existence of a giant local void. Note, however (see below) that this assumed local void does not exist on our past light cone but in hypersurfaces $t =$ constant and it is thus in a space-like relation to us.

The arguments brought in defense of the constant $t_B$ assumption are only expectations that should not be treated as objective truth unless they are verified by calculations. Such calculations have already been done, and it turns out that the inhomogeneities in $t_B$ needed to explain a structure of present radius 30 Mpc is of the order of a few hundred years and that this age difference between the oldest and youngest region would generate CMB temperature fluctuations equal to $\Delta T/T = 3.44 \times 10^{-6}$ and $\Delta T / T = -2.35 \times 10^{-6}$, respectively \cite{KB09}. This is well-hidden in the observational errors at the current level of precision. Therefore, there is no justification to the assumption $t_B =$ constant.

Hence, to determine the $\Lambda = 0$ L--T model best fitting the observations the two independent functions of the radial coordinate defining this model must be left totally arbitrary. We thus impose neither an hand-picked algebraic form to these functions, nor $t_B$ = constant \cite{CBKH09}.

Using a set of data -- the angular diameter distance together with the mass density in redshift space -- both assumed to have the same form on our past light cone as in the $\Lambda$CDM model, the two left arbitrary L--T functions are determined and give the mass distribution in spacetime.

\begin{figure}
  \includegraphics[height=.3\textheight]{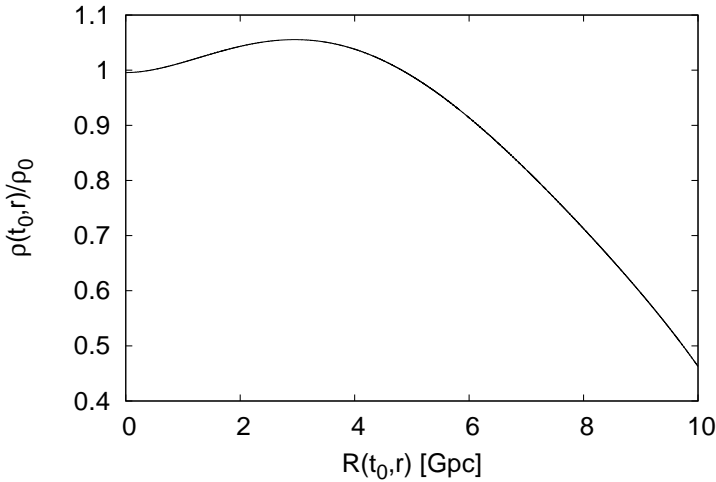}
  \caption{The hump: the density at the present instant as a function of the current areal radius.}
\end{figure}

In this case, the current density profile does not exhibit a giant void but a giant hump and we are located in a shallow and wide basin on top of this hump (FIGURE 5).

Note, however, that as the giant void, the giant hump is not directly observable. It exists in the space $t =$ now, of events simultaneous with our present instant in the cosmological synchronization, i.e. it is in a space-like relation to us. However, contrary to the giant void of the GBH type which does not reproduce the $\Lambda$CDM $\rho(z)$ function and can thus be expected to be testable by high redshift galaxy counts, the giant hump is not observable in $\rho(z)$ or in the number count data, since, by construction, it is designed to reproduce them faithfully.

Why such a difference between the density distribution on our past light cone and in the $t =$ now space? It is due to one basic feature of the L--T model (and in fact of all inhomogeneous models): on any initial
data hypersurface, whether it is a light cone or a $t =$ constant space, {\em the density and velocity distributions are two algebraically independent functions of the radial position}. Thus the density on a later hypersurface may be quite different, since it depends on both initial functions. Whatever initial density distribution can be completely transformed by the velocity distribution. For example, as predicted by Mustapha and Hellaby \cite{MH01} and explicitly demonstrated by Krasi\'nski and Hellaby \cite{KH04}, any initial overdensity can evolve into a void and vice versa. In FLRW models, there are no physical functions of position, and all worldlines evolve together. Thus, while
dealing with an L--T (or any inhomogeneous) model, one must forget all
Robertson--Walker-inspired prejudices and expectations.

It is also worth emphasizing that the existence of the giant hump is most probably a feature of the particular L--T models that we ended up with. This result must not become the starting point of a new paradigm in observational cosmology, aimed at detecting the hump. Before this happens, it must be decided at the theoretical level whether the hump is a necessary implication of L--T models properly fitted to other observations.

\subsection{An example of L--T Swiss-cheese: Marra et al.'s model}

The model is a lattice of L--T bubbles with radius $\geq$350 Mpc in an Einstein-de Sitter (EdS) background. Initially, the void at the center of each hole is dominated by negative curvature and a compensating overdensity matches smoothly the density and curvature EdS values at the border of the hole \cite{MK07}.

Since the voids expand faster than the cheese, the overdense regions contract and become thin shells at the borders of the bubbles while underdense regions turn into emptier voids, eventually occupying most of the volume.

This shows once more that the evolution of the voids bends the photon paths and affects more photon physics than the geometry of the inhomogeneities. But, here, the inhomogeneities of the model are only able to partly mimic the effect of dark energy. These results have induced us to consider QSS Swiss-cheese models which exhibit enhanced structure evolution (see the subsection `Structure evolution with quasi-spherical Szekeres models') and with which we thus hope to increase the fit to the `dark energy' component \cite{BC09}.

\section{Extracting the cosmic metric from observations}

The inverse problem of deriving the arbitrary functions of a L--T model from observations is very much involved. This is the reason why most of the authors who have tried to deal with this issue have added some a priori constraints to the model \cite{CR06,TN07}.

However, Lu and Hellaby \cite{LH07}, McClure and Hellaby \cite{MCH07} and Hellaby and Alfedeed \cite{HA08} have initiated a program to extract the metric from a set of observations. This is the full inverse problem and it is not degenerate. To date it has assumed the metric has the L--T form, as a relatively simple case to start from, though the long term intention is to remove the assumption of spherical symmetry.

These authors have developed and coded an algorithm that generates the L--T metric functions, given observational data on the redshifts, apparent luminosities or angular diameters, number counts of galaxies, estimates for the absolute luminosities or true diameters and source masses, as functions of $z$. This allows both of the physical functions of an L--T model to be determined without any a priori assumption on their form.

\section{Conclusion}

The increasing precision of observational data implies that FLRW models must now be considered just a zeroth order approximation, and linear perturbation theory a first order approximation whose domain of validity is an early, nearly homogeneous Universe.

In the nonlinear regime, which was entered since structures formed, there is no escape from the use of exact methods (or of averaging schemes aiming at investigating this issue from the standpoint of backreaction).

In the era of `precision cosmology', the effect of the inhomogeneities on the determination of the cosmological models cannot be ignored. Inhomogeneous models constitute an exact perturbation of the Friedmann background and can reproduce it as a limit with any precision. This is the reason why they are fully adapted for the purpose of studying astrophysical and cosmological effects and for constructing precise models of universe.

While using L--T models with a central observer to represent our `local' Universe averaged over angles around us, a giant void is not mandatory to explain away dark energy. A giant overdensity can also do the job. However while neither the void nor the overdensity are directly observable, the giant void alone can be tested with observations of the density function on our past light cone. The giant hump will need more and more precise data to be constrained.

Exact inhomogeneous solutions can be employed not only for studying the geometry and dynamics of the Universe, but also to investigate the formation and the evolution of structures. They give enhanced formation efficiency and might therefore help solving the problem of structure formation pertaining to the standard model.

While the L--T models have been mostly used up to now for modeling the inhomogeneities of the Universe, the need of getting rid of spherical symmetry for this purpose will lead the cosmological community to consider other solutions, and among them QSS models, for the future developments of inhomogeneous cosmology.

\end{document}